\documentclass[preprint,12pt]{elsarticle}

\usepackage{amssymb}

\usepackage{graphicx}
\usepackage{amsmath}
\usepackage{amsfonts}

\usepackage{caption}
\usepackage{subcaption}

\journal{Artificial Intelligence in Medicine}

\begin{document}

\begin{frontmatter}



\title{Assessment and treatment of visuospatial neglect using active learning with Gaussian processes regression}


\author[inst1]{Ivan De Boi}

\affiliation[inst1]{organization={Faculty of Applied Engineering, Department Electromechanics, Research Group InViLab (http://www.invilab.be), University of Antwerp},
            addressline={Groenenborgerlaan 171}, 
            city={Antwerp},
            postcode={2020}, 
            country={Belgium}}

\author[inst2]{Elissa Embrechts}
\author[inst2]{Quirine Schatteman}
\author[inst1]{Rudi Penne}
\author[inst2]{Steven Truijen}
\author[inst2]{Wim Saeys}

\affiliation[inst2]{organization={Department of Rehabilitation Sciences and Physiotherapy, Research Group MOVANT, University of Antwerp},
            addressline={Universiteitsplein 1}, 
            city={Wilrijk},
            postcode={2610}, 
            country={Belgium}}

\begin{abstract}
Visuospatial neglect is a disorder characterised by impaired awareness for visual stimuli located in regions of space and frames of reference. It is often associated with stroke. Patients can struggle with all aspects of daily living and community participation. Assessment methods are limited and show several shortcomings, considering they are mainly performed on paper and do not implement the complexity of daily life. Similarly, treatment options are sparse and often show only small improvements.

We present an artificial intelligence solution designed to accurately assess a patient's visuospatial neglect in a three-dimensional setting. We implement an active learning method based on Gaussian process regression to reduce the effort it takes a patient to undergo an assessment. Furthermore, we describe how this model can be utilised in patient oriented treatment and how this opens the way to gamification, tele-rehabilitation and personalised healthcare, providing a promising avenue for improving patient engagement and rehabilitation outcomes.

To validate our assessment module, we conducted clinical trials involving patients in a real-world setting. We compared the results obtained using our AI-based assessment with the widely used conventional visuospatial neglect tests currently employed in clinical practice. The validation process serves to establish the accuracy and reliability of our model, confirming its potential as a valuable tool for diagnosing and monitoring visuospatial neglect. Our VR application proves to be more sensitive, while intra-rater reliability remains high.

\end{abstract}

\begin{graphicalabstract}
\includegraphics[width=\textwidth]{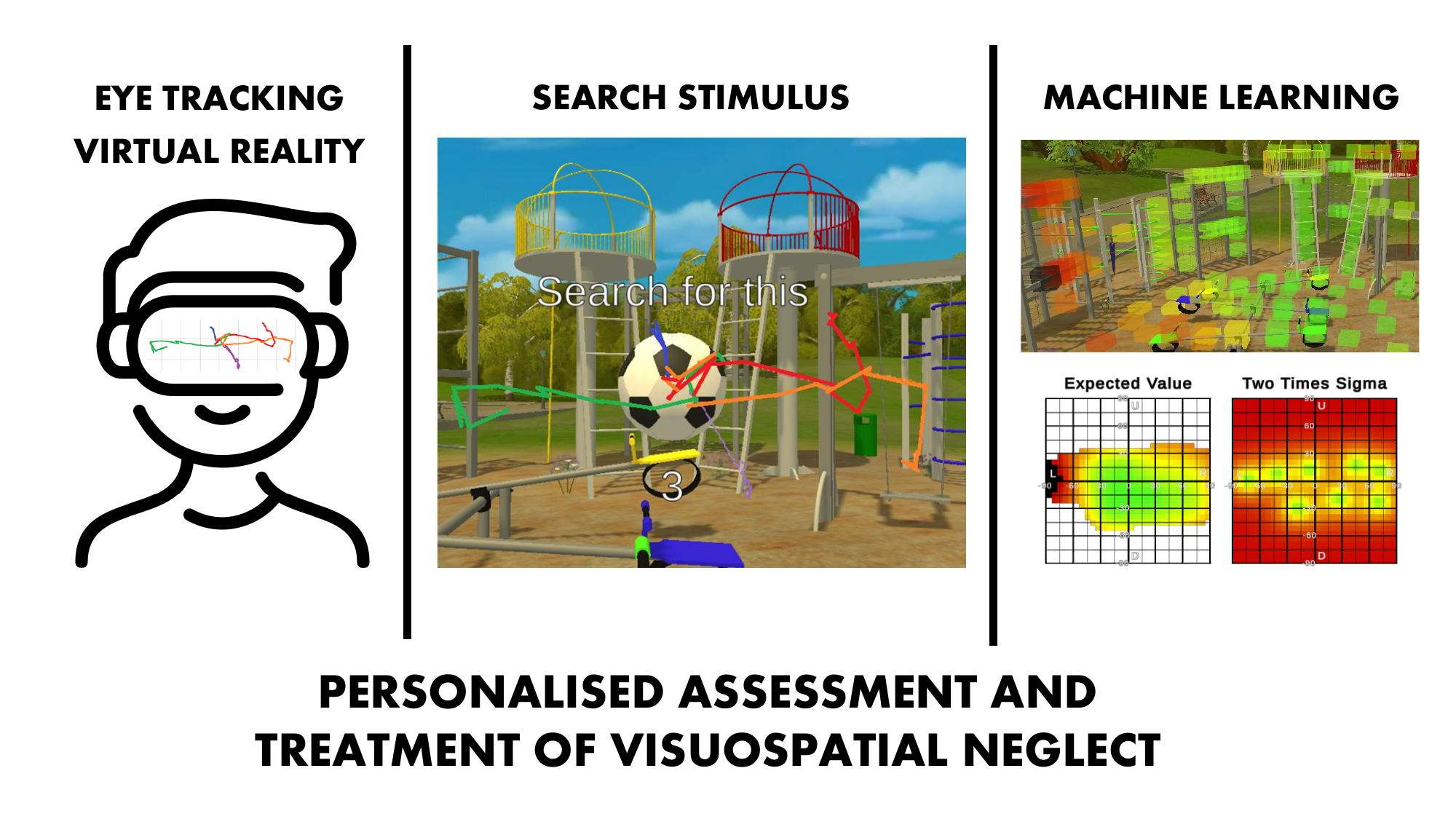}
\end{graphicalabstract}

\begin{highlights}
  \item We formulate an artificial intelligence framework designed to assess search behaviour and attentional deficits in a virtual 3D setting.
  \item We implement an active learning method based on Gaussian process regression to reduce the effort it takes a patient to undergo this assessment. 
  \item We describe how this model can be utilised in treatment and how this opens the way to gamification, tele-rehabilitation and personalised healthcare.
  \item We describe our VR application which demonstrates these practises.
  \item We validate our findings on patients in a clinical setting. We compare them to the standard tests used in practice today.
\end{highlights}


\end{frontmatter}


\section{Introduction}

Visuospatial spatial neglect (VSN) is a cognitive disorder characterised by a lateralized attention deficit with reduced attention towards the contralesional hemispace and increased capture of information in the ipsilesional hemispace \cite{Zebhauser, Heilman1979, Fellrath2012, Stigchel2010}. VSN is a heterogeneous disorder that can vary both in regions of space and in frames of reference. It is often one of the deficits associated with stroke \cite{Nijboer2013_2}. Approximately half of stroke patients worldwide experience VSN within the first two weeks of onset, and at least 40 percent of patients still experience symptoms a year post-stroke \cite{Nijboer2013}. VSN has a significant impact on postural control, mobility, independence during daily-life and community participation after stroke \cite{Bosma2020,Embrechts2021}. 

Accurately diagnosing VSN in clinical practice is thus highly relevant. However, current assessment methods for VSN are limited in this ability and show several shortcomings, including a lack of ecological validity, reliability, and distinctiveness between the different types  (e.g., peri-personal, extra-personal) and severities of VSN \cite{Dawson2017, Donovan2007, Barrett2019}. Moreover, patients are often suffering from fatigue and a decreased attention span. Similarly, the current treatment options are sparse, while the effects are often small \cite{Barrett2006}. We elaborate on this in depth in Section \ref{related_work}.

As a result, patients remain highly dependent on the spontaneous recovery of the neural system because of the small treatment effects currently seen in clinical practice. However, a huge number of patients will have persistent VSN after rehabilitation leading to substantial loss of community participation with a high dependency on (in)formal care. 

Therefore, the overall primary objective of this work is to deliver a solution for some of the aforementioned problems by means of an artificial intelligence (AI) based virtual reality (VR) application. AI allows us to assess and treat VSN in a "smart" way, while VR has the potential to provide a three-dimensional simulation of a real-life environment increasing ecological validity. In short, we place stimuli in different virtual environments (both peri-personal and extra-personal) and measure the time it takes a patient to find them. From these measurements, we construct a heatmap of the patient's VSN. The application tracks both eye- and head movements to gather data about visual scanning strategies and spatial attentional deficits.

Our assessment module is smart, in the sense that VSN  will be mapped efficiently and accurately using AI. This effectively lowers the necessary amount of stimuli that has to be placed in order to acquire a viable assessment of a patient's VSN, resulting in shorter assessment sessions. In practice, patients suffering from VSN and possibly other pathologies as well, are in no condition to spot dozens of stimuli in one time consuming session. Therefore, we rely on AI techniques that can handle small datasets. In this work, we implement active learning \cite{Pasolli2011}. Starting with only a few measurements, the next location to place a stimulus in the virtual world is chosen on a criterion that maximises the gain in information. This results in reducing the overall number of measurements needed compared to measuring everywhere in a grid. We base our method on Gaussian processes \cite{Rasmussen2006}. This probabilistic machine learning technique makes predictions that are accompanied by an uncertainty. The latter can be exploited. A more detailed explanation is given in Section \ref{methods}.

Once the precise VSN characteristics of a certain patient is captured by our AI model, we can apply this in our treatment module. Exercises will be provided based on the results of the assessment module enhancing individualised specific treatment. More details on this treatment module are given in Section \ref{treatment}.

Besides the technical benefits of using both AI and VR described above, there is an added benefit. VR is a mature technology that requires only commercially available hardware that is reasonably priced. The latest generation of head mounted displays (HMD) are stand alone, meaning they do not require a powerful desktop PC or laptop and are cordless. This facilitates the adaptation by acute hospitals, rehabilitation centres, private practises, and even home users. The latter opens the possibility for tele-rehabilitation. Our application allows for independent training of the AI model by the patient at home under supervision of a remote therapist.

The contributions of this work include:
\begin{itemize}
  \item  We formulate an artificial intelligence framework designed to assess search behaviour and attentional deficits in a virtual 3D setting.
  \item We implement an active learning method based on Gaussian process regression to reduce the effort it takes a patient to undergo this assessment. 
  \item We describe how this model can be utilised in treatment and how this opens the way to gamification, tele-rehabilitation and personalised healthcare.
  \item We describe our VR application which demonstrates these practises.
  \item We validate our findings on patients in a clinical setting. We compare them to the standard tests used in practice today.
\end{itemize}

The rest of this paper is structured as follows. In the next section we reference related work. In Section \ref{methods}, we give some theoretical background on Gaussian processes and active learning. Section \ref{ours} describes how we implemented this in assessment and treatment. We present our results in Section \ref{results} and we discuss these in \ref{discussion}. Finally, conclusions are provided.

\section{Related work} \label{related_work}

\subsection{Visuospatial neglect}

In this section we describe the condition known as visuospatial neglect and reference to existing work on its assessment and treatment. We also explain the shortcomings of said methods that are still used in practice today.

The medical condition “stroke” is a condition defined by the World Health Organisation by rapidly developing symptoms and signs of a focal (or at times global) neurological impairment, lasting more than 24 hours or leading to death, with no apparent cause other than that of vascular origin. The number of stroke events in Europe is projected to rise from 1.1 million in 2000 to 1.5 million per year by 2025, largely due to the ageing population. The high burden on the community is the result of stroke being the leading cause of complex disability
worldwide in adults (WHO). These people are often suffering from sensorimotor impairments in both the upper and lower limbs and the trunk. Furthermore, stroke survivors can have speech and comprehension problems, and cognitive impairments such as disorientation in time and space, decreased information processing time and volume, memory problems and attentional deficits. One of these post-stroke attentional deficits is visuospatial neglect (VSN) characterised by impaired awareness for visual stimuli located on the contralesional side of space. In other words, patients do not pay attention to stimuli on the side opposite to the brain lesion (in hemispheric strokes). This results in problems with reporting, responding or orienting towards contralesional visual stimuli, which cannot be attributed to sensory or motor impairments alone. This means that VSN is located at the processing level of information rather than impairments at the input level (e.g., eyes, vestibular apparatus). 

Furthermore, it is a heterogeneous disorder as it can vary in regions of space such as peri-personal (within reach) or extra-personal (far) space and frames of reference such as egocentric (first-person view) or allocentric (object-based) reference frames. Within the first two weeks after stroke onset, VSN occurs in approximately 50 percent of patients worldwide \cite{Nijboer2013}.

Spontaneous neurological recovery of VSN follows a natural logistic pattern of improvement in some patients, depending on severity and type of VSN, within the first 12 to 14 weeks post-stroke. Afterwards, the curve flattens and severity of VSN remains almost invariant, leaving at least 40 percent of patients with initial VSN still with symptoms a year post-stroke \cite{Nijboer2013}. For Europe alone, this means that approximately 300,000 new stroke survivors each year encounter persistent VSN. This is important since people with VSN experience significant postural impairments and a high fall risk. In addition, consequences can be more practical of nature as patients with attentional deficits are, for example, unaware of the traffic lights at street crossings or even traffic in general, but also lack the ability to find products at grocery stores. It is obvious that people dealing with cognitive impairments encounter difficulties in all aspects of ADL (Activities of Daily Living) and community participation, and can even lack the ability to live independently at home \cite{Nijboer2014, Nijboer2013_2, Bosma2020, Barrett2014}. With this in mind, research on VSN is crucial to improve assessment and explore new treatment options.

Although research in cognitive impairments is advancing, there are still huge gaps in the literature considering assessment and treatment of attentional deficits such as VSN. Currently, VSN assessment usually consists of pen-and-paper tasks that are administered in a quiet room where distractions are minimal. Although these tests are easy to administer and assess underlying cognitive impairments such as VSN, they suffer from several shortcomings.

First, research has reported a huge lack of ecological validity \cite{Dawson2017}. Performances on pen-and-paper tests do not correlate well to ADL, which results in a poor understanding of the difficulties patients encounter in daily life \cite{Donovan2007}. Furthermore, these tests cannot differentiate well between different types of VSN and the exact severity of the deficit.

Second, standard assessment is sub-optimal since VSN is a clear three-dimensional problem and current assessment methods are two-dimensional, therefore lacking indeed ecological validity, reliability and
distinctiveness between different types of VSN (e.g., space within reach versus far space). Moreover,
it has been stated that many patients with VSN are not even identified as a neglect patient and
therefore not treated accordingly \cite{Barrett2019}.

Third, in standard assessment, we do not have accurate information on eye and head movements which are important to assess the strategies or compensations the patient uses to complete the task.  The information on both eye- and head movements will be imperative to have a good understanding of the (compensatory) strategies used by the patient.  Furthermore, poor understanding of the severity and nature of the VSN derived from the standard assessment methods highly impacts treatment efficacy.
Treatment options are sparse  and effects are often limited.   In  literature,  especially  visual  scanning training,  active  limb  activation,  prism  adaptation training, and sustained attention training are the better options. However, these  treatments  are  often  delivered  in  clinical,  controlled  settings,  which lack the complexity and stimuli of real-life environments.  Bringing patients within such complex environments during rehabilitation (e.g., a leisure park or shopping centre) is clinically not feasible due to associated health risks and costs for the health care system.  The use of 3D VR may overcome these issues, and may deliver ecologically valid, patient-centred training whilst still being in a controlled and safe environment \cite{Salatino2023}.

\subsection{AI and VR aided assessment}

In this section we zoom in on methods described in the literature that assess visuospatial neglect and are based on AI and VR.

In \cite{dvorkin2012mapping}, a subject was presented 3D spheres in a 3D virtual environment projected on a large screen. With the push of a response button, the subject could indicate the perceptance of the stimulus. Head movement was tracked with a special rolling shutter HMD. Linear regression models were implemented to map VSN. In comparison, our Gaussian process model allows for non-linear regression.  

The work of \cite{Jang2016} describes field of regard, field of view and attention bias measurement based on VR. Again, only head movement is measured while subjects search for visual stimuli in a virtual environment.

A fully developed VR game to treat stroke-induced attention deficits is described in \cite{Huygelier2017}. They implemented a dynamic difficulty adjustment (DDA) mechanic that tailors the experience to the needs of the patient. This serves two purposes. First, this allows the game to present high priority stimuli more frequently in the neglected field than in the good field. Second, this algorithm adjusts the difficulty of the game to an appropriate level for each player. Their DDA is based on a 2D Gaussian distribution, of which the mean is initialised at the centre of the visual field and is adjusted based on the median locations of missed targets at a fixed rate. This approach differs from our method, which is based in a Gaussian process, not a Gaussian distribution. Confusingly, the names for these are quite similar. In our work, the Gaussian process provides a Gaussian distribution for every possible stimulus location, not just for the overall field of view. A more thorough explanation is given in Section \ref{methods}. Another example of an immersive virtual reality game to train spatial attention orientation after stroke, can be found in the pilot study given in \cite{Huygelier2020}.

In the above-mentioned work, no active learning strategies were implemented to reduce the number of stimuli placement iterations. The economic validity was not taken into account, except in the VR games. Moreover, no eye-tracking solutions were developed.

\section{Methods} \label{methods}

\subsection{Gaussian processes}
In this section, we provide a brief overview of Gaussian processes. A more in-depth explanation can be found in \cite{Williams2006}. This book defines a Gaussian process (GP) as a continuous collection of random variables, any finite subset of which is normally distributed as a multivariate distribution.

In this work, we describe a dataset of $n$ observations as $\{(\mathbf{x}_{i},y_{i})\}^{n}_{i=1}$, where $\mathbf{x}$ is an input vector of dimension $d$ and $y$ is a scalar-valued measurement. The process of regression is to find a function $f:\mathbb{R}^d\rightarrow\mathbb{R}$,
\begin{equation}\label{GPR}
	y = f(\boldsymbol{\mathbf{x}})+\epsilon, \quad \epsilon\sim\mathcal{N}(0,\sigma_{\epsilon}^2) ,
\end{equation}
with $\epsilon$ being identically distributed observation noise. This function can be sampled from a GP, which can also be defined by its mean $m(\boldsymbol{\mathbf{x}})$ and \textit{covariance function} $k(\boldsymbol{\mathbf{x}},\boldsymbol{\mathbf{x}}')$, also denoted as
\begin{equation}
f(\boldsymbol{\mathbf{x}})\sim\mathcal{GP}(m(\boldsymbol{\mathbf{x}}),k(\boldsymbol{\mathbf{x}},\boldsymbol{\mathbf{x}}')) . 
\end{equation}

A covariance function is parametrized by a set of hyperparameters $\boldsymbol{\theta}$. Training a Gaussian process means finding an optimal set of values for those hyperparameters. This is achieved by maximising the log marginal likelihood. In this work we use BFGS, a quasi-Newton method described in \cite{Liu1989}. A commonly used covariance function is the squared exponential kernel (SE). It is applicable in a wide variety of applications because it generates infinitely differentiable (smooth) functions. It has the form:
\begin{equation} 	 	
	k_{SE}(\mathbf{x}, \mathbf{x}') = \sigma^2_{f}\exp \left( -\frac{\left|\mathbf{x}-\mathbf{x}' \right|^2 }{2l^2}\right) ,
\end{equation}
in which $\sigma^2_{f}$ is a height-scale factor and $l$ the length-scale that determines the radius of influence of the training points.

In our work, the inputs of the Gaussian process are points in the 2D space that represents a person's ideal field of view (without neglect). The outputs for this model are the recorded times it takes to find a particular stimulus. Neglected areas are given a maximum (truncated) value. We thus learn the mapping, or function, from a point in a person's field of view to search times.

\begin{figure}
     \centering
     \begin{subfigure}[b]{0.45\textwidth}
         \centering
         \includegraphics[width=\textwidth]{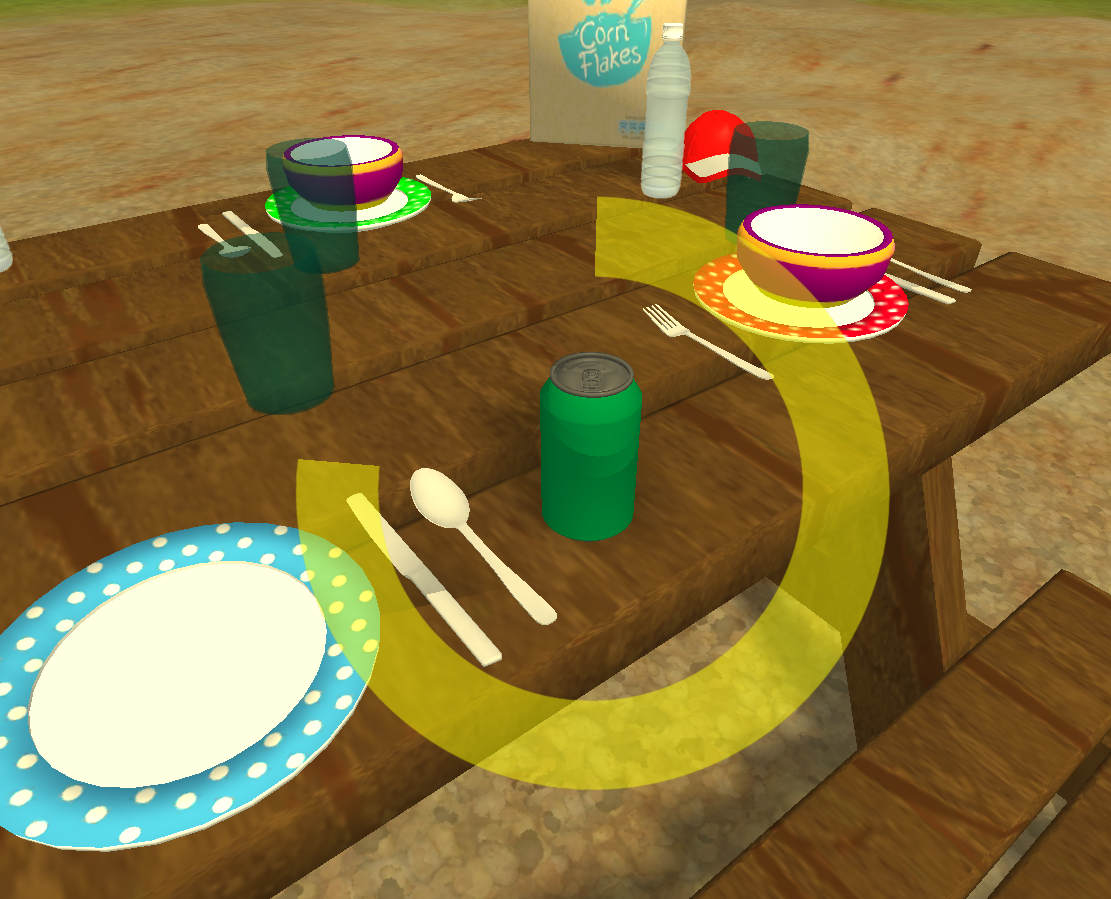}
         \caption{}
         \label{}
     \end{subfigure}
     \begin{subfigure}[b]{0.45\textwidth}
         \centering
         \includegraphics[width=\textwidth]{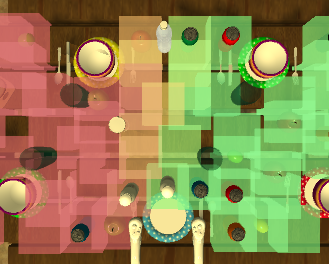}
         \caption{}
         \label{}
     \end{subfigure}     
     
    \begin{subfigure}[b]{0.45\textwidth}
         \centering
         \includegraphics[width=\textwidth]{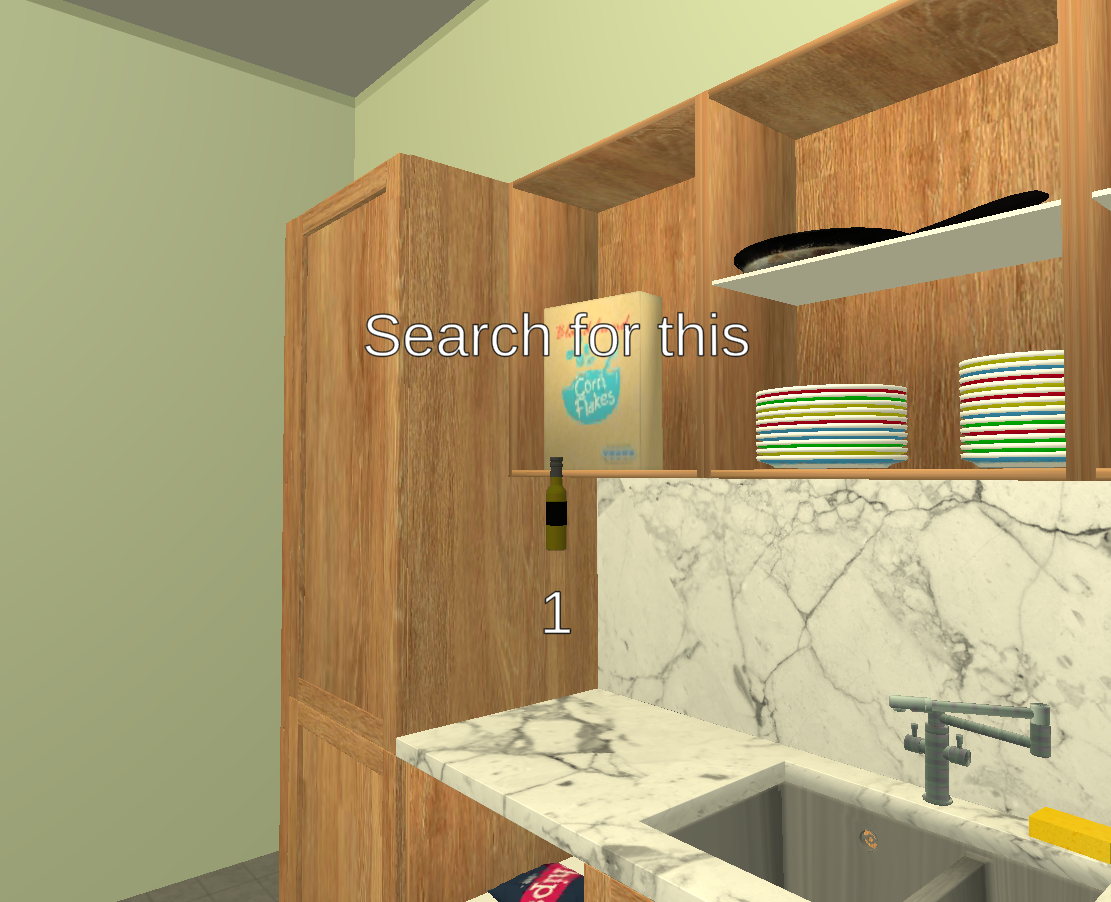}
         \caption{}
         \label{}
     \end{subfigure}
     \begin{subfigure}[b]{0.45\textwidth}
         \centering
         \includegraphics[width=\textwidth]{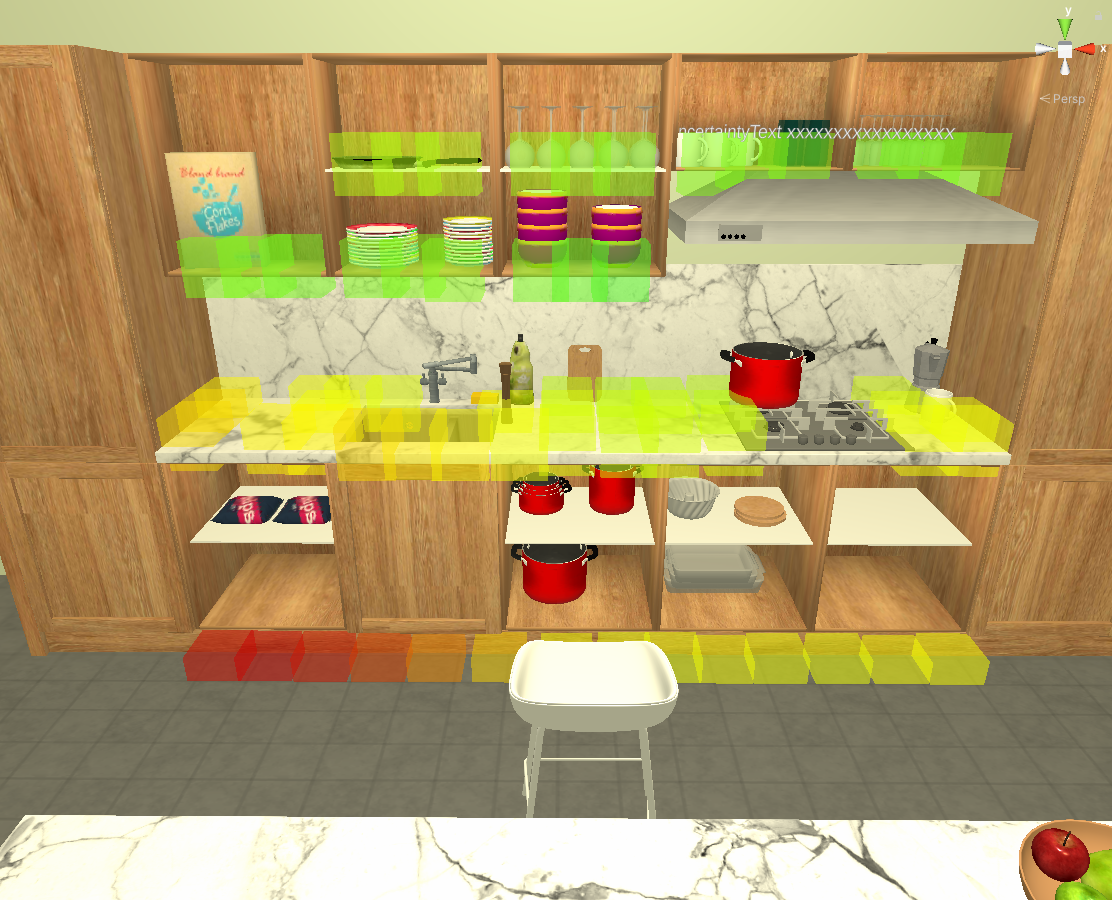}
         \caption{}
         \label{}
     \end{subfigure}     
     
    \begin{subfigure}[b]{0.45\textwidth}
         \centering
         \includegraphics[width=\textwidth]{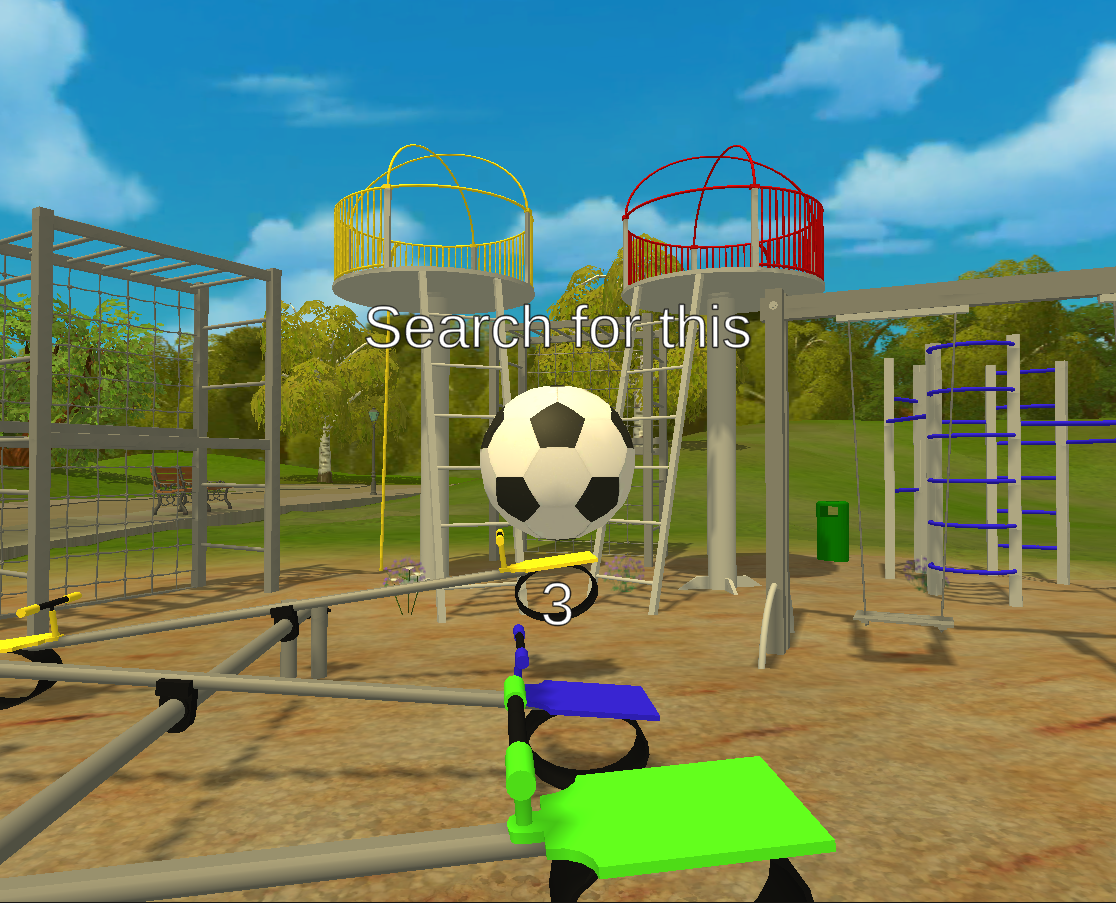}
         \caption{}
         \label{}
     \end{subfigure}
     \begin{subfigure}[b]{0.45\textwidth}
         \centering
         \includegraphics[width=\textwidth]{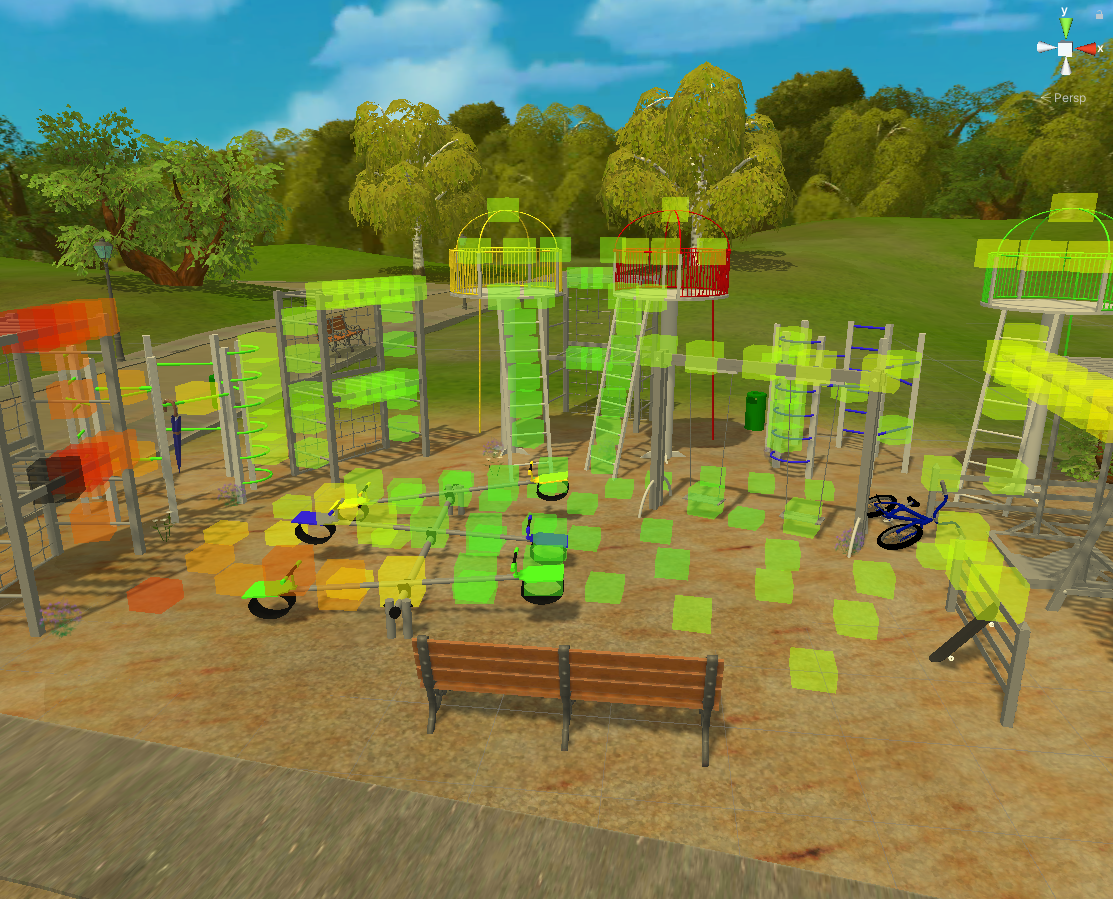}
         \caption{}
         \label{}
     \end{subfigure}
     
    \caption{Screenshots capturing the view as seen through the HDM for the table (a), kitchen (c) and playground environment (e). Visualisations for the therapist showing a 3D view of a patient's neglect for those same environments (b, d, f). A green colour for the cubes represents short search times. A red colour indicates long search times. A black cube in the left part of (f) means a location where a stimulus was never found.}   
    \label{fig:screenshots}
\end{figure}

We can visualise this 2D function as a heat map, as shown in Figure \ref{fig:screenshots}. A Gaussian process provides two values for each 2D coordinate in a patient's field of view: a mean (expected value) and an uncertainty (variance or two times sigma). We plot the mean on the left heat map and the uncertainty on the right. The axes represent the left-right and up-down direction in the field of view, expressed in degrees. The green colour in the mean heat map results from stimuli being found fast. The red and black are areas where stimuli are either not found or found only after a significant amount of time. As for the two times sigma plot, the green areas are locations where the uncertainty is low. This means stimuli have been placed there. Red areas are regions where no measurement has taken place and thus where uncertainty is high.

\subsection{Active learning}

The process of mapping a patient's VSN accurately is time consuming. It depends on both the measurement of the search time for a given stimulus and the number of stimuli we want to present to a patient. The latter defines the resolution of the VSN situational map. The strategy to overcome this is to train a machine learning algorithm to predict these search time values for every possible point in a patient's field of view. The goal now is to train the model as accurately as possible given only a limited number of data points.

Our underlying AI model is a Gaussian process. It serves as a surrogate for the true VSN situation we wish to learn. By that, we mean the mapping from 2D points in a person's field of view to search times for those points. By placing stimuli at specific locations in a 3D virtual environment which correspond to 2D points in a person's field of view, we can make point measurements of this search time value for that specific point. In our application, we work with predefined locations, instead of a continuous range. We refer to these locations as spawn points.

Our active learning strategy consists of the following steps:
\begin{enumerate}
\item We start by presenting stimuli one by one at a small amount of spawn points. These are sampled at random from the list of all possible spawn points. Alternatively, those points could lay on a grid, be Latin hypercube sampled or chosen from a Sobol sequence \cite{Noe2019,Fang2005,Jones1998}.
\item A Gaussian process is trained on this initial small dataset.
\item The spawn point from the 2D field of view with the highest uncertainty (variance) in the GP's posterior distribution is chosen to place a stimulus next. This method is called Uncertainty Sampling (US). Alternatively, the spawn point which reduces the total variance of the posterior could be chosen. This method is called Integrated Variance Reduction (IVR). 
\item Step 3 is repeated until a certain criterion is met. When limited by the patient's condition, this could be a fixed number of iterations. Another criterion is convergence in the posterior distribution, which means that adding new data points no longer has a significant result on the predictions of the GP.
\end{enumerate}

These steps are what is known as active learning by the machine learning community \citep{Williams2006, Buisson2020, Pasolli2011, Gessner2020,Oladyshkin2020}. It is active in the sense that the learning is based iteratively on what is measured. This is in contrast to learning, or training, on an entire pre-measured dataset. 

All computations have to be performed in real time. Longer computation times could result in lag between a patient's head movement and what can be seen through the HMD or a drop in the frame rate. This would significantly add to the risk of nausea for the patient. We implemented Uncertainty Sampling because it is faster to compute \cite{Sacks1989}.

\section{Our application} \label{ours}

Our application is made with the game engine Unity (version 2020.2.5f1). We build this for the HMD Pico Neo 2 Eye, which has six degrees of freedom, a 4k RGB display, a 101° field of view and built-in Tobii eye tracking hard- and software. We implemented a VSN assessment module based on active learning and a treatment module that provides the patient with a personalised experience. Stimuli are placed on spawn points picked by the AI model. A patient has to gaze at that stimulus for a certain amount of time in order to mark it as detected. During which, a yellow circle fills up around the stimulus, as shown in Figure \ref{fig:screenshots}.a. 

\begin{figure}
     \centering     
     \begin{subfigure}[b]{0.6\textwidth}
         \centering
         \includegraphics[width=\textwidth]{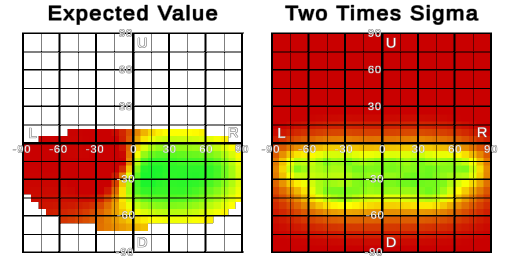}
         \caption{}
         \label{}
     \end{subfigure}
     
    \vfill
     \begin{subfigure}[b]{0.6\textwidth}
         \centering
         \includegraphics[width=\textwidth]{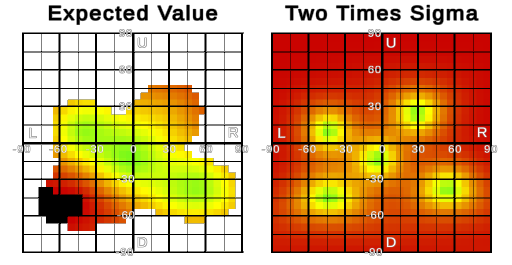}
         \caption{}
         \label{}
     \end{subfigure}
     
    \vfill
     \begin{subfigure}[b]{0.6\textwidth}
         \centering
         \includegraphics[width=\textwidth]{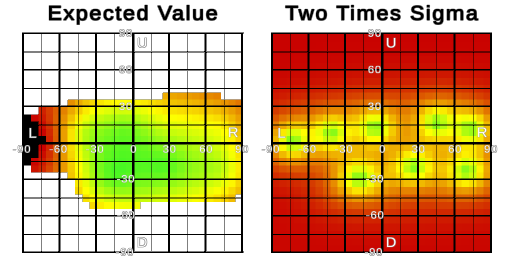}
         \caption{}
         \label{}
     \end{subfigure}    
    \caption{Heat maps depicting the VSN for a patient's field of view corresponding to the situations depicted in Figure \ref{fig:screenshots} b, d and f respectively. These images are generated on the HMD itself. The resolution is kept low for performance reasons. Regions where the uncertainty is large (red in the right plot) are made white in the left plot.}   
    \label{fig:heatmaps}
\end{figure}

\begin{figure}
     \centering
     \begin{subfigure}[b]{0.45\textwidth}
         \centering
         \includegraphics[width=\textwidth]{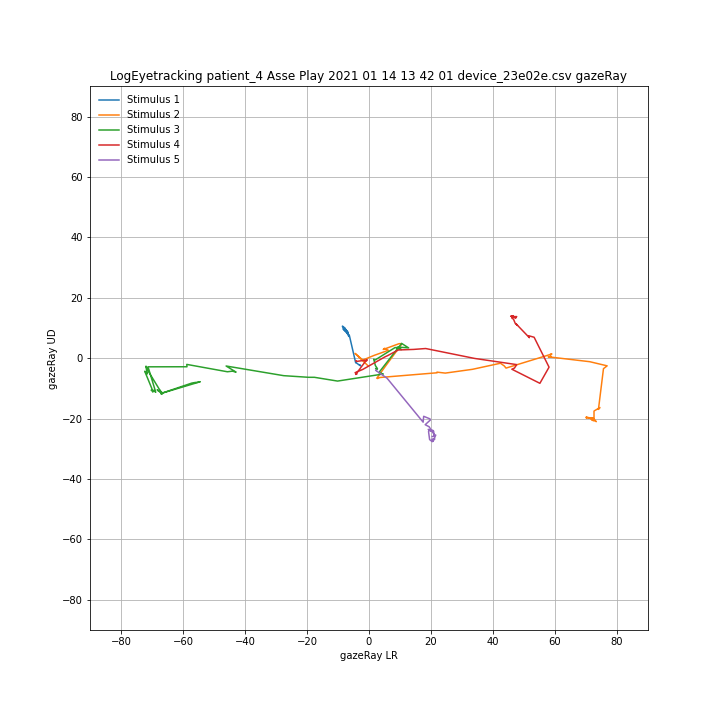}
         \caption{}
         \label{}
     \end{subfigure}
     \begin{subfigure}[b]{0.45\textwidth}
         \centering
         \includegraphics[width=\textwidth]{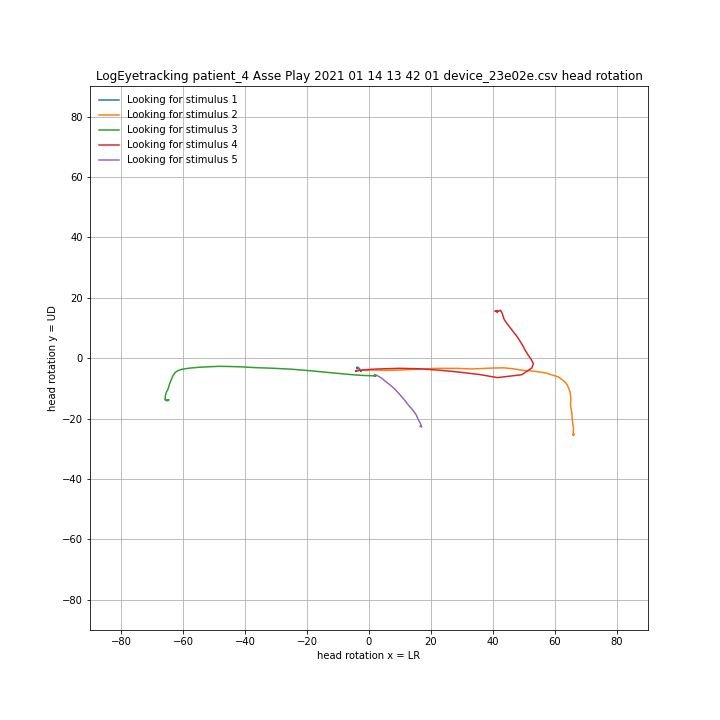}
         \caption{}
         \label{}
     \end{subfigure}
     \begin{subfigure}[b]{0.45\textwidth}
         \centering
         \includegraphics[width=\textwidth]{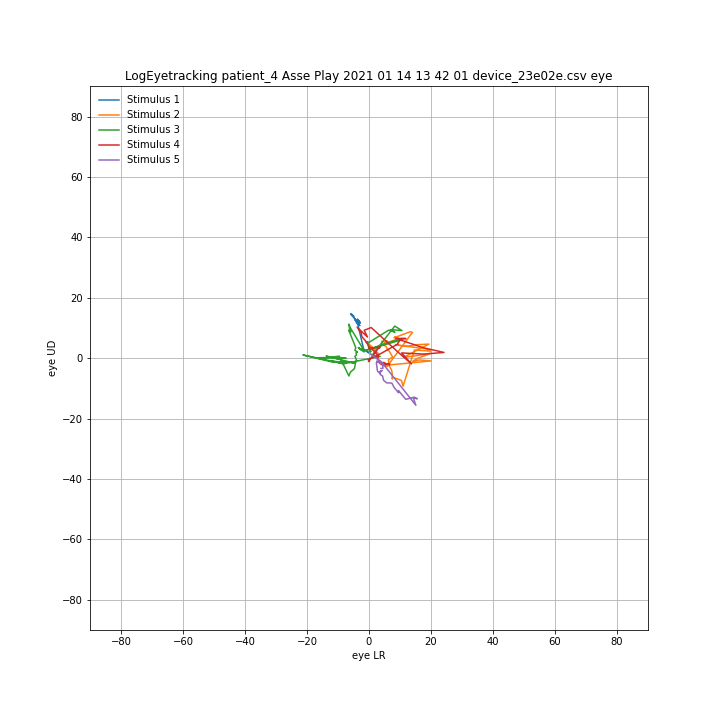}
         \caption{}
         \label{}
     \end{subfigure} 
    \caption{Movement of a person's gaze in the 3D world (a), decomposed in his head movement (b) and his eye movement (c).}   
    \label{fig:gaze}
\end{figure}

\subsection{Assessment}\label{assessment}

In our application, we implemented three different scenes: a picnic table, a kitchen and a playground. This allows for the placement of stimuli in three different regions. These regions correspond to three clinically significant distances: near peripersonal (reaching) space, far peripersonal space and extrapersonal (far) space. The 3D effect of a patient's VSN can thus be studied more accurately. Screenshots are given in Figure \ref{fig:screenshots}.

A therapist has access to the following parameters to tailor the application to a patient's needs: game modus (assessment or treatment), scene (table, kitchen, playground), patient's number, number of stimuli to show, maximum allowed search time, minimum gaze distance (in degrees) and difficulty level. The latter controls the amount of distractors in the scene. These are static objects that resemble stimuli to be found. For example, a plate on the table. We conveniently grouped some of these parameters into editable presets.

The results of our assessment module are presented in heatmaps, as shown in Figure \ref{fig:heatmaps}. In our case, these results are the predictions for search times predicted by the machine learning algorithm. We implemented a Gaussian process as a surrogate model. This allows us to perform active learning. Predictions of a Gaussian process for a search time at a specific point in a person's field of view are in fact Gaussian distributions. They consist of a mean or expected value and a variance, here reformulated as a two times sigma value. The left part of the heatmap shows the expected value for a point in the field of view of a person. A green colour on the heatmap indicates a small search time, red indicates a longer time and black means the stimulus was not found. The right part of the heatmap shows the two times sigma values that accompany the expected values. The same colour scheme is kept, meaning that the green areas are regions where the visuospatial neglect is well mapped, while the red areas are regions where there is still some uncertainty. These red areas are candidates to put future stimuli in the active learning process. A red point in the expected value heatmap that correlates with a green point in the two times sigma heatmap indicates that in this area we can be sure there is an attention deficit. On the other hand, a green point in the expected values heatmap that correlates with a red point in the two times sigma heatmap, indicates we are not sure there is no neglect in this area.

\subsection{Treatment} \label{treatment}

Once a patient's VSN situation has been learned by the AI model, it can be exploited in therapy. Our application saves the trained Gaussian process models for each assessment session. During treatment, this model can be queried for spawn points in the vicinity of the border between the neglected areas and the non-neglected area. Placing stimuli in random places, which are either in the neglected area or not, results in exercises that are either too hard or too easy. Moreover, systematically placing stimuli on this border can train a patient to seek out this region in his field of view. The aim of this therapy, also known as cueing \cite{genannt2023combined}, is to gradually shift that border. 

\section{Results} \label{results}

To validate the methods used in our assessment, we conducted a case-control cross-sectional pilot study. It has been approved by the local Ethics Committee of the Antwerp University Hospital (Belgian registration number: B3002020000216). Data collection took place between October 2020 and September 2021 at the Rehabilitation Hospital RevArte in Antwerp, Belgium. All participants were asked to give signed informed consent.

We worked with three groups of participants: healthy controls (HC), stroke participants without VSN (N-) and stroke participants with VSN (N+). The number of participants required for this study was calculated in advance to ensure sufficient statistical power. A sample size of twenty participants in each of the three groups was necessary to achieve an 80\% chance (power=0.80) of detecting a 5\% difference between groups on the gaze ray area. Due to of the Covid-19 pandemic, only thirty-eight participants could be recruited. Participants were recruited from the stroke population of the rehabilitation hospital Revarte, Antwerp, Belgium, which is not an acute stroke unit. Participants between 16 and 99 years old with an ischemic or haemorrhagic stroke in the right hemisphere were eligible for inclusion. Subjects were excluded if they fit one of the following criteria: unable to sit in a wheelchair or chair, unable to understand the procedure due to cognitive impairment, having impaired eyesight (i.e., visual field deficit), and refusal to participate. Cybersickness symptoms were assessed with the Simulator Sickness Questionnaire \cite{Kennedy1993}, as well as user experience.

Participants were seated in a wheelchair or on a straight back chair. Their trunk was restricted to the chair. Prior to the assessment, the screen of the head mounted display was projected onto a laptop and the eye tracking was calibrated to the participants' eye positions by the investigator. A tutorial of the far space version (the playground) was shown in which participants were instructed to look for the objects shown in the search task and hold their gaze until a yellow circle was completed. The investigator examined if participants had enough range of motion in the neck to view all objects. Participants were allowed to move their head and eyes in every direction without moving their trunk.

The levels for peripersonal (near) and for extrapersonal (far) neglect were completed in one session by all participants. This means the table and playground environment. All participants were tested again on the third difficulty level of both versions within one week to investigate the intra-rater reliability.

Cognitive function was tested with the mini mental state examination (MMSE). For testing neglect, the following three tests were administered: broken hearts test (BHT), line bisection test (LBT) and visual search time test (VSTT). These tests are commonly used in practice and serve as the standard to which we will compare our VR application. The first test was administered on paper and the other two on a Wacom tablet (Saitama, Japan). All tests were administered in the same order and before assessing neglect using our VR application. Healthy subjects were only tested with the VR application. All participants were tested again on the third difficulty level in both near and far version within one week to be able to evaluate the intra-rater reliability of the VR application. During the VR assessment, the eye and head movements were tracked ten times per second. To assess neglect in the VR application, the maximal rightward angle was compared with the maximal leftward angle on three levels: gaze ray (GR), head movement (HR) and eye movement (ER). The maximum angles were added to form the search area middle (SAM) which should be close to 0. A SAM higher than 0 would mean the participant searched further to the right than to the left, a value lower than 0 would mean the participant searched further to the left than to the right. The SAM GR refers to the comparison of the maximal left and right gaze ray position, the SAM HR refers to the comparison of the maximal left and right head rotation angles and SAM ER refers to the comparison of the maximal left and right eye rotation angles. Due to common misunderstandings, the first level of each version in the VR programme was used as a trial. Therefore only the second and third level of the table and the playground version were used in the statistics. The levels are referred to as T2 and T3 in the second and third near space version, respectively, P2 and P3 is used to refer to the second and third level of the far space version, respectively. 

Statistical tests were conducted with IBM SPSS Statistics, version 21. Descriptive statistics were performed for the collected variables of the participants. To evaluate whether or not this application could detect neglect, a ROC curve was conducted to establish a cut off value. To evaluate the intra-rater reliability, maximal angles and the SAM of gaze ray, head rotation and eye rotation were checked on normality with Shapiro Wilk. Normally distributed parameters were checked on significant differences with a paired t-test. Not normally-distributed parameters were checked with a Wilcoxon signed rank test. To evaluate differences between groups, the maximal angles and the SAM of gaze ray, head rotation and eye rotation were compared with the Kruskal Wallis.

Thirty-eight participants completed the VR programme. It was stated that a stroke patient had VSN if 2 out of 3 neglect test (BHT, LBT, VSTT) were positive. Group 0 (HC) consisted of 15 participants (five female, mean age 55.4y, SD 24.4) , group 1 (N-) consisted of thirteen participants (6 female, mean age 63.2y, SD 18.2) and group 2 (N+) consisted of ten participants (six female, mean age 67y, SD 19.3). To verify whether the VR application can detect neglect, a ROC curve was conducted for all of the SAM's. Four of the parameters showed an area under the curve that was higher than 80\%. In the case of the SAM GR P2, it could be assumed that from a difference of 0.00585 or higher in maximal left and right gaze ray, a patient has neglect. In this case 80\% of N+ group are diagnosed correctly and eight out of twelve N- participants seem to show neglect as well (participants 1,10,22,25,27,30,31). Looking at the SAM GR P3, it could be assumed that from a difference of 0.0218 or higher a patient shows neglect. However, here we only find three out of twelve N- participants to have neglect, whereas only 70\% of the N+ is diagnosed correctly (participants 16,25,31). In the case of the SAM HR P3, it could be assumed that from a difference of 7.4° or higher in maximal left and right head rotation a participants shows neglect. Here, nine out of thirteen show neglect (participants 1,2,15,16,17,22,25,26,31). Two participants (25 and 31) test positive on all three cut off scores for extrapersonal neglect. This could imply that the VR application is more sensitive than the standard tests used in practice today (BHT, LBT and VSTT). Looking at the SAM ER T2, it could be assumed that from a difference of 3.96° in maximal eye rotation, four out of twelve participants have neglect. Here 80\% of the neglect patparticipantsients are diagnosed correctly and 7\% percent of the healthy subjects are false positive.

To verify whether the VR application had good intra-rater reliability, the SAM and the maximal angles of GR, HR and ER of the first assessment were compared to the second assessment, which was conducted within one week after the first. The Shapiro-Wilk was used to check normality. A paired t-test was conducted for the five parameters that showed a normal distribution (GR T3 max, HR T3 min and max, SAM HR T3, HR P3 min). None of these parameters showed significant differences between the first and second assessment of the VR application.

For the other parameters a Wilcoxon Signed Rank test was used (GR T3 min, SAM GR T3, GR P3 min and max, SAM GR P3, HR P3 min, SAM HR P3). Only two parameters showed significant differences. A Wilcoxon Signed Rank test indicated that the SAM HR P3 was lower in the first testing (mean rank = 8) than in the retesting (mean rank = 13.11), Z=-2.981, P=0.003. This means that participants rotated their heads more to the left in the first than in the second test session. A Wilcoxon Signed Rank test indicated that the maximal GR angle in the third playground level was lower in the first testing (mean rank = 7.13) than in the retesting (mean rank = 13.38), Z=-2.034, P =0.042. This means participants seemed to look further to the right in the second testing session than in the first. This could be explained by a small learning effect. However, there was no significant difference on the SAM GR P3 (Z=-1.851,P=0.064). Only two parameters showed significant differences. All ten other parameters seemed to be comparable, indicating that the VR application might have a good intra-rater reliability.

The Cybersickness Questionnaire showed that no participants experienced moderate to high levels of cybersickness symptoms after completing the VR application.

\section{Discussion} \label{discussion}

Our VR application seems to be more sensitive than the pen and paper tests. By taking wider angles into account, the application can better imitate real life and reflect the impact of neglect on daily life more closely. According to the maximal search angles and the search area middle, there was only little difference between the initial testing and the re-testing, which indicates that our VR application is reliable. However, these results need to be interpreted with caution, as a larger study population is needed. Moreover, the treatment module was not included in the clinical investigation.

The AI constructs a personalised heatmap, which is a valuable tool for clinicians as it provides a clear view of a patient's personal VSN. This will make it easier to interpret the condition and explain it to not only the patient itself, but also friends, family and other caregivers. This can help to create awareness in the patient's surroundings.

Our application also keeps detailed log files of the tracking of both eye and head movement. As shown in an example in Figure \ref{fig:gaze}, the gaze of a patient in the 3D world can be decomposed in head and eye movement. We can see that for this individual searching for stimuli, the eyes remain scanning around the centre, while the head looks more left and right. This information can be used in research on search strategies, neck movement etc. This application can also be beneficial to assess and treat other attention-based conditions, search strategies and neck movements. This would be a fruitful area for further work. 

We plan to further expand the gamification aspect by implementing a game in the kitchen scene where the player has to follow a certain recipe.

\section{Conclusion}

In this paper, we described an artificial intelligence based virtual reality application that is able to assess and treat a patient's visuospatial neglect. By placing stimuli at certain spawn points in a patient's field of view and measuring the search times for those, we can effectively map his VSN situation using a surrogate model. In our work this is a Gaussian process. Moreover, this allows for active learning, which effectively lowers the number of stimuli to be placed. We perform measurements in the field of view where the uncertainty is largest. This results in a more feasible (shorter) experience for patients while still maintaining high accuracy.

We developed our application using the Unity 3D game engine. It works on a Pico Neo 2 Eye, which has built-in eye tracking. As an added bonus, this provides therapists and clinical researchers with detailed data on eye and head movements that can be used in research on search strategies etc.

We compared our work to standard tests used in practice today: broken hearts test, line bisection test and visual search time test. Our VR application proves to be more sensitive, while intra-rater reliability remains high.

Moreover, the trained AI models reveal the border between neglected areas and non-neglected areas in a patient's field of view. This can be exploited in therapy, where stimuli can be placed around this border in order to gradually move it. Our AI implementation tailors both the assessment and the treatment of visuospatial neglect to the patient itself.

Finally, this treatment could be performed without the intervention of a therapist, opening the way for tele-rehabilitation. Although in practice, for most patients, handling the headset and starting the application still requires assistance from a caregiver. More usability research is needed to make this as practicable as possible.

\section*{Declaration of competing interest}
The authors declare that they have no known competing financial interests or personal relationships that could have appeared to influence the work reported in this paper.


\section*{Patent}
The method in this work is protected under patent PCT/EP2021/084817.

 \bibliographystyle{elsarticle-num} 
 \bibliography{cas-refs}





\end{document}